\documentclass[10pt,aps,prd,twocolumn,nofootinbib,floatfix]{revtex4}
\usepackage{amsfonts,amsmath,amssymb,amsthm}
\usepackage{graphicx}
\usepackage{color}

\usepackage{enumerate}

\begin{document}

\author{Edward Malec}\email{edward.malec@uj.edu.pl}
\affiliation{Institute of Theoretical Physics, Jagiellonian University, {\L}ojasiewicza 11, 30-348 Krak\'{o}w, Poland}
 
\title{The mass density contrast in perturbed    Friedman-Lemaitre-Robertson-Walker cosmologies }

\begin{abstract}
 We analyze the evolution of the mass density contrast in spherical perturbations of flat Friedman-Lemaitre-Robertson-Walker cosmologies. Both dark matter and dark energy are included. In the absence of dark energy the evolution equation coincides with that obtained by Bonnor within the ``Newtonian cosmology''.  
\end{abstract}
\maketitle

\section{Introduction}

We shall analyze the evolution of perturbations of flat FLRW spacetimes using the $1+3$ splitting of the spacetime. 
The original aim of this paper was just to find the general relativistic version of the well known result of Bonnor \cite{Bonnor}, assuming isothermal perturbations and using the comoving coordinates.
The    main conclusion concerning the temporal behaviour of the mass density contrast --- in the absence of dark energy --- coincides with that of Bonnor and  also with a later analysis of  \cite{Martel}), for   perturbations comoving with the background matter. The case of the nonzero cosmological constant was  not  investigated by Bonnor. In such a case the evolution equation for the mass density contrast   differs from that found earlier by Martel \cite{Martel}.

\section{Selfgravitating fluids within spherically symmetric spacetimes }

We shall assume only spherical symmetry, without  spatial homogeneity. Some of the resulting Einstein equations   had been found   by Lemaitre in 1930's   \cite{Lemaitre1931, Lemaitre}, who studied stability of Einstein static universes. Tolman and Bondi extended results of Lemaitre for a selfgravitating dust   \cite{Tolman,Bondi1947}. The resulting class of metrics is often referred to as the Lemaitre-Tolman-Bondi spacetimes.  In 1960's Misner and Sharp \cite{Misner}, and Podurets  \cite{Podurets} again analyzed these equations, but in the case of perfect  gas;  they  extended  in particular the Lemaitre-Tolman-Bondi  concept of the quasilocal material mass. Its expression will be given below.  

We assume the Einstein equations $R_\mathrm{\mu \nu }-g_\mathrm{\mu \nu }R=8\pi T_\mathrm{\mu \nu }-\Lambda g_{\mu \nu }$, where the  stress-energy tensor is defined as $T_\mathrm{\mu \nu }=\left( \varrho +p\right) U_\mu U\nu +pg_{\mu \nu }$ and $\Lambda $ is the cosmological constant. The coordinate $4-$velocity is normalized, $U_\mu U^\mu=-1$. Here $\varrho $ and $p$ denote the mass density and pressure, respectively.

We shall assume that we are given a $1+3$ foliation, with foliation leaves characterized by constant time, $t=const$. 
The line element is taken in the form
\begin{equation}
ds^2=-N^2dt^2+\hat a dr^2+R^2\left( d\theta^2+\sin^2 \theta d\phi^2\right),
\label{1}
\end{equation}
where  the radius $0\le r <\infty $ and the angular variables satisfy $0\le \phi <2\pi $, $-\pi /2 \le \theta \le \pi/2$. The lapse $N$ and the areal radius $R$ depend on time $t$ and the coordinate radius $r$. We adopt the standard condition that the speed of light $c$ and the gravitational constant $G$  are equal to unity.

This metric is diagonal, so that we shall calculate  extrinsic curvatures from the formula  $K_{ij}=\frac{1}{2N} \partial_tg_{ij}$ \cite{MTW}.    The condition of isotropy implies that two of them are equal,  $\mathrm{K_{\phi  }^{\phi  }}=\mathrm{ K_{\theta   }^{\theta  }}$. The nonzero components 
of $K_{ij}$ read  
\begin{eqnarray}
\mathrm{trK}&=&{\partial_t(\sqrt{\hat a}R^2) \over N \sqrt{\hat a} R^2},~~~
\mathrm{K_r^r}={1\over 2N \hat a}\partial_t\hat a,~~~   \mathrm{K_{\phi  }^{\phi  }}=\nonumber\\
&&
\mathrm{ K_{\theta   }^{\theta  }}={\partial_tR \over N R}={1\over 2}(\mathrm{trK}-\mathrm{K_r^r})
\label{IV.2.1}
\end{eqnarray}
Usually one assumes that coordinates are comoving. We shall impose  a foliation condition as in  the standard $1+3$ formulations of Einstein equations, by putting a condition  onto extrinsic curvatures of leaves of a foliation.  We shall assume the following    
\begin{equation}
\Delta (R(r,t),t) = ({R(\mathrm{trK}-\mathrm{K_r^r})\over 2})^2
\label{VI.2.1}
\end{equation}
where $\Delta $ is defined as  \cite{Malec1999}:
\begin{eqnarray}
\Delta (R(r,t),t)&=&{-3\over 4R}\int_0^R \tilde R^2(\mathrm{K_r^r})^2d\tilde R  + {1\over 4R}
\int_0^R\tilde R^2(\mathrm{ tr K })^2d\tilde R+\nonumber\\
&&{1\over 2R}\int_0^R \mathrm{trK K}_r^r\tilde R^2
d\tilde R.
\label{VI.1.2}
\end{eqnarray}
Differentiation of both sides of   Eq. (\ref{VI.1.2})  with respect the coordinate radius $r$ yields, using the momentum constraint of Einstein equations \cite{MTW}   and the definition of the mean curvature $\hat p=2
\partial_r\ln 
R/\sqrt{\hat a}$ \cite{Malec1999}, 
\begin{equation}
R\mathrm{(trK-K_r^r) }{16\pi j_\mathrm{r}R\over \hat   p}=0.
\label{VI.2.2}
\end{equation}
Herein we define $j_\mathrm{r}=NT^0\mathrm{r}/\sqrt{\hat a}$.

This implies that fluids are comoving in chosen coordinates, 
\begin{equation}
j_\mathrm{r}=0,
\label{VI.2.3}
\end{equation}
provided that there are no minimal surfaces,   $\hat p\ne 0$ and  $\mathrm{tr K}\ne K_\mathrm{r}^\mathrm{r}$. On the other hand, it appears that in comoving coordinates  $ \mathrm{tr K}= \partial_R\left( R^3\mathrm{(trK-K_r^r) }\right)/(2R^2)$    (see  Sec. IV A). The areal velocity 
$R\mathrm{(trK-K_r^r)}/2$ constitutes a part of the initial data of Einstein equations --- see the forthcoming equation (\ref{IV.2.5aaa}). Thus under the conditions $\hat p\ne 0$ and  $\mathrm{tr K}\ne K_\mathrm{r}^\mathrm{r}$ our foliation equation 
(\ref{VI.2.1}) is equivalent to the standard assumption of comoving coordinates.

Notice that   now the material energy-momentum tensor reads  $T_0^0=-\varrho $, $j_\mathrm{r}=0$ and   $T_\mathrm{r}^\mathrm{r}=p=T_\mathrm{{\theta }}^\mathrm{{\theta }}$; we deal with perfect fluids. The cosmological constant is responsible for   the dark energy $\varrho_\Lambda$  and pressure $p_\Lambda$ contributions:
\begin{equation}
\varrho_\Lambda =\frac{\Lambda}{8\pi},
~~~p_\Lambda =- \frac{\Lambda}{8\pi}.
\end{equation}
 In such a case the quasilocal mass of Misner and Sharp \cite{Misner}, and Podurets  \cite{Podurets}, contained in a coordinate sphere of a radius $r$,  is given by the formula
\begin{equation}
m(R(r))=2\pi \int_0^r\tilde R^3\hat p\sqrt{a} \left(  \rho +\varrho_\Lambda \right)dr.
\label{VI.2.4}
\end{equation}
For the sake of concise notation we shall define 
\begin{equation}
U(r)=\frac{R(r)}{2}\left(\mathrm{tr K(r)}- K(r)_\mathrm{r}^\mathrm{r}\right);
\label{U}
\end{equation}
this quantity represents areal velocity of a comoving particle of gas, $U=\partial_0R/N$.  
The mean curvature $\hat p$ of centered  spheres can be calculated to be \cite{Malec1999}
\begin{equation}
\hat p= \frac{2}{R(r)}\sqrt{1-\frac{2m(R(r))}{R(r)}+U^2(r)}.
\label{hatp}
\end{equation}
One can show that the mass  defined in (\ref{VI.2.4})  changes as follows \cite{Misner}
\begin{equation}
\partial_t m(R(r)) =
-4\pi \left[N R^2 U\left( p +p_\Lambda \right)\right](r).
\label{VI.2.5}
\end{equation}
Moreover, by direct calculation one gets from (\ref{VI.2.4})
\begin{equation}
{\partial_rm(R)\over \sqrt{\hat a}}= 2\pi R^3\hat p\varrho .
\label{VI.2.6}
\end{equation}
These equations  should be supplemented by two conservation  equations  
\begin{equation}
  N  \partial_r p
 +  \partial_rN (p+\varrho)=0,
\label{VI.2.7}
\end{equation}
and
\begin{equation}
\partial_t  \varrho=
 -N \mathrm{tr K} (p +\rho ).
\label{VI.2.8}
\end{equation}
The Einstein evolution equations  reduce to the single equation   
\begin{eqnarray} 
\partial_tU=-\frac{m(r)}{R^2}-4\pi \left( p+p_\Lambda \right) RN+\frac{\hat pR}{2\sqrt{\hat a}}\partial_rN. 
\label{IV.2.5aaa}
\end{eqnarray}
 
\section{The Friedman type solution}

Assuming that matter consists of dust and imposing in addition homogeneity    on   slices of constant time $t$,
one gets  from equations (\ref{VI.2.4} --- \ref{IV.2.5aaa}) the Friedman  metric   $ds^2=-dt^2+a^2\left( dr^2+r^2d\Omega^2\right)$. Thus the lapse $N=1$. The conformal factor $a(t)$ satisfies Friedman equations:
\begin{eqnarray}
&&\varrho_0+\varrho_\Lambda=\frac{3}{8\pi}H^2\nonumber\\
&&-\frac{dH}{dt}=4\pi \varrho_0  \nonumber\\
&&\frac{d\varrho_0}{dt}=-3H\varrho_0.
\label{Friedman}
\end{eqnarray}
(Only two of the three equations are independent.)

The extrinsic curvatures of this solution are equal to the Hubble parametr $H\equiv \frac{da}{adt}$,
\begin{equation}
K_\mathrm{r}^\mathrm{r} =K_\mathrm{\theta}^\mathrm{\theta}=K_\mathrm{\phi}^\mathrm{\phi}=H,
\end{equation} while its trace is $\mathrm{tr }K=3H$.
The velocity $U$ reads now $U=HR$. The mean curvature of centered 2-spheres within the $t=const$ slice is now the same as in the flat space: $\hat p=2/R$.
 
  This solution describes   a flat, homogeneous and isotropic universe filled with comoving dust of the density $\varrho_0$, that is expanding with the Hubble recession velocity $H=\dot a/a$. The product $\varrho_0a^3$ is constant in time.

\section{Evolution of small spherical inhomogeneities in a   FLRW universe}

We assume that the  background (Friedman-type) universe is dotted by isolated, locally isotropic mass density perturbations $\delta\varrho$, so that the   mass density is split into the background part $\varrho_0$ and the perturbation $\delta\varrho$: $\varrho=\varrho_0+\delta\varrho$. The mass perturbations are isothermal --- they exert pressure $p=c^2_s \delta\varrho$. The metric of the perturbed spacetime reads $ds^2=-N^2dt^2+\hat adr^2+R^2d\Omega^2$; we use comoving coordinates. Far from these perturbations the lapse $N$ tends to 1 and the spatial part of the metric is approaching the background metric  $a^2\left( dr^2+r^2d\Omega^2\right)$. We assume --- similarly as Bonnor in his analysis of \cite{Bonnor} ---   that this perturbing isothermal gas is comoving with the background dust. (Let us remark, that perturbations do not have to comove with the background dust --- see a different scenario discussed in \cite{Pelykh}.)  For the matter of convenience we shall locate   our coordinate system in the symmetry  center of a perturbation.  

The areal velocity $U=\partial_0R/N=R(\mathrm{tr K}-\mathrm{K^r_r})/2$ is split into the background and perturbed parts as follows
\begin{equation}
U=H(t)R+\delta_U,
\label{edeta1} 
\end{equation}
where $H(t)$ is the Hubble constant at the time  $t$.

We need initial data  --- for the areal velocity  $U=\partial_0R/N$ and the mass density  $\varrho$ --- for the two evolution equations (\ref{VI.2.8}) and  (\ref{IV.2.5aaa}). They  are defined as follows at an initial hypersurface labelled by the world time $t_0$.  The initial value of the perturbing component $\delta_U$ is small but otherwise it is a free datum.
The initial mass density $\varrho $ is given as the sum of the background mass density $\varrho_0$ at the time $t_0$ and the small initial perturbation   $\delta\varrho$, with the condition that far from the center $\varrho$ approaches $\varrho_0(t_0)$.

The main aim of forthcoming calculation is the derivation of the wave equation that rules the evolution of the mass density contrast $\delta \varrho /\varrho_0$. We shall get also an evolution equations for the velocity perturbation $\delta_U$.

\subsection{The extrinsic curvature}

The first part of the calculation is actually exact --- we do not  need the assumption of small perturbations in order to get the trace of the extrinsic curvature
  \begin{equation}
 \mathrm{tr K}=\frac{\partial_R\left( R^2U \right)}{ R^2  }
 \label{edeta2} 
 \end{equation}
 of hypersurfaces of constant world time $t$.
 
 Formula (\ref{edeta2}) is valid in all slicings of  spherically symmetric spacetimes  cosmological models that   asymptotically coincide with flat slicings of cosmological flat FLRW models. We allow for dark energy (cosmological constant) and various forms of {\it comoving} matter --- dust and  fluids.This formula is known (see for instance \cite{Malec1999,Mach}), but we derive it here for the sake of completeness.

  We have from the definition of extrinsic curvatures 
 \begin{equation}
 \mathrm{tr K}=\frac{\partial_0\left( R^2\sqrt{\hat a}\right)}{NR^2\sqrt{\hat a}}.
 \label{edeta3} 
 \end{equation}
The  quantity $\sqrt{\hat a}$ in the nominator of (\ref{edeta3}) can be replaced by 
 \begin{equation}
 \sqrt{\hat a} =\frac{2\partial_rR }{ \hat  p R};
 \label{edeta4} 
\end{equation}
here $\hat p$ is the mean curvature of the coordinate sphere $r=const$.  Thus (\ref{edeta3}) yields
\begin{equation}
 \mathrm{tr K} = 2\frac{\partial_0  R  }{NR } +
 \frac{2\partial_0 \partial_rR}{N\hat pR\sqrt{\hat a}}+\frac{\hat pR}{N\sqrt{\hat a}}\partial_0\frac{ 1 }{\hat pR}.  
 \label{edeta5} 
\end{equation}
The first term is just $   2U/R$. Changing the order of differentiation, we can write the second term   as 
$$\frac{2\partial_r  \left( \frac{\partial_0R}{N}N\right)}{N\hat pR\sqrt{\hat a}}= 2\frac{\partial_rU}{\hat pR\sqrt{\hat a}}+2U\frac{\partial_rN}{N\hat pR\sqrt{\hat a}}. $$
 Replace now the coordinate radius $r$ by the areal radius $R$ and notice that $\frac{2\partial_r}{\hat pR}=\partial_R$. We obtain   the following form of the  second term of (\ref{edeta5}): $$\frac{2\partial_0 \partial_rR}{N\hat pR}=\partial_RU +U \frac{\partial_RN}{N}. $$ The calculation of the third term in 
(\ref{edeta5}) is a little bit longer. Recall (see formula (\ref {hatp}))  that the mean curvature  $\hat pR=2\sqrt{1-\frac{2m(R(r,t),t)}{R(r,t)}+U^2(r,t)}$. Its differentiation with respect   time yields, after using the mass conservation equation (\ref{VI.2.5}) and the   Einstein equation describing the evolution of $U=R(\mathrm{tr-K^r_r})/2$ (see  equation   (\ref{IV.2.5aaa})):
\begin{equation}
\partial_t\frac{2}{\hat pR}=-\frac{2U}{\hat pR}\partial_RN
\label{edeta6}
\end{equation}
Combining the three terms of (\ref{edeta5}), we arrive at the formula (\ref{edeta2}).

  In the case of  small spherically symmetric perturbations we can   use     the splitting (\ref{edeta1}) of the radial velocity.  We immediately arrive at the following corollary.

{\bf Conclusion}. Assume a perturbed FLRW flat universe. The trace of the extrinsic curvature of constant time  hypersurfaces,  in the foliation defined by the assumption of comoving particles, is given by
\begin{equation}
\mathrm{trK}=3H+\frac{\partial_R\left( R^2\delta_U\right)}{R^2}.
\label{edeta6}
\end{equation}

{\it Remark. The alternative way to derive the formula (\ref{edeta2}) is to write down the momentum constraint (i.e., the Einstein equation $R_{0i}-\frac{R}{2}g_{0i}=8\pi T_{0i}$), using the metric (\ref{1}). The $r$-component of the constraint can be expressed as   (\ref{edeta2}), in comoving coordinates.}

 \subsection{The lapse}
 
In what follows we need the lapse function $N$; it can be obtained from (\ref{VI.2.7}). We assumed that the pressure is isothermal  in perturbed FLRW universes,  $p=c^2_s \delta \varrho =c^2_s   \varrho_0 \delta$, where we introduced the mass density contrast
\begin{equation}
\delta \equiv \frac{\delta \varrho}{\varrho_0}.
\label{edeta6a}
\end{equation} 
If the mass density contrast is small, $\delta \ll 1$, then (\ref{VI.2.7}) yields $\partial_rN\approx -c^2_s\partial_r\delta $. Far from the center $N\rightarrow 1$; thus 
\begin{equation}
N\approx 1-c^2_s\delta .
\label{edeta7}
\end{equation}

This implies that the time derivative of the areal radius evolves as
\begin{eqnarray}
\partial_0R&=&UN\approx \left(H R+\delta_U\right)\left( 1- c^2_s \delta\right) \approx \nonumber\\
&&H R+\delta_U - c^2_s \delta  H R.
\label{edeta8}
\end{eqnarray}

\subsection{Evolution of the mass density contrast}

We investigate   perturbations of (flat)  FLRW universes with dust (including dark matter) and dark energy. Let us summarise   the relevant information. The material pressure $p_0=0$ and the sum of the background energy density satisfies  $\varrho_0+\varrho_\Lambda=\frac{3H^2}{8\pi}$. The metric scale factor $a(t) $ of the background metric  can be obtained  from equations (\ref{Friedman}).   The lapse up to the first perturbation is given by (\ref{edeta7}) and Eq. (\ref{edeta8}) reads now  $\partial_0R\approx HR+\delta_U - c^2_s HR$.

Equation (\ref{IV.2.5aaa}) can be written as 
\begin{equation}
\partial_0U=-\frac{m(R)}{R^2}N -4\pi \left( c^2_sR  \varrho_0\delta+p_\Lambda \right) RN+ \frac{\hat p^2R^2}{4}\partial_RN .
\label{edeta9}
\end{equation}
 Zeroth order terms (see Section III) drop out. Thus the linear perturbations satisfy the equation
\begin{equation}
\frac{1}{a}\partial_0\left(a \delta_U\right)=-\frac{\delta m(R)}{R^2} -  c^2_s \partial_R\delta   .
\label{edeta10}
\end{equation}
We employed   (\ref{edeta1}) and   (\ref{edeta7}) --- (\ref{edeta9}) in the process of deriving (\ref{edeta10}).

One can show that in the leading order of $O(\delta)$ the following rule holds 
\begin{equation}
\partial_0\partial_R\left(R^2\delta_U\right)=\partial_R\left( \frac{R^2}{a}\partial_0\left( a\delta_U\right) \right).
\label{edeta9a}
\end{equation}
The mass density conservation equation is given by (\ref{VI.2.8}). Using the derived earlier expressions for the lapse $N$ and the trace of the extrinsic curvature $\mathrm{tr K}$, we get 
\begin{equation}
\partial_0 \delta \varrho +3H\delta \varrho +\frac{\varrho_0}{R^2}\partial_R\left( R^2\delta_U\right)= 0.
\label{edeta11}
\end{equation}
Dividing both sides by $\varrho_0$, we obtain
\begin{equation}
\partial_0 \delta      +\frac{1}{R^2}\partial_R\left( R^2\delta_ U\right)= 0.
\label{edeta12}
\end{equation}
Differentiate now both sides of (\ref{edeta12}) with respect time, use formula (\ref{edeta9a})  and equation (\ref{edeta10}). After straightforward calculation we arrive at 
\begin{equation}
\partial^2_0 \delta -\frac{c^2_s}{R^2}\partial_R\left( R^2\partial_R\delta \right) -\frac{3}{2}H^2 \delta +2H \partial_0\delta = 0.
\label{edeta13}
\end{equation}
 Notice also that Eq. (\ref{edeta13}) is a wave equation --- thus it possesses a kind of travelling wave pulses that move within the coordinate sphere that encloses the perturbed   initial data.
 
Equation (\ref{edeta13}) is  equivalent to the corresponding  Bonnor equation describing the evolution of the mass density contrast \cite{Bonnor, Martel} when the cosmological constant is absent.    In order to see this, perform  the Fourier transformation  of (\ref{edeta13}) and insert $H^2 =8\pi \varrho_0/3$. Then  one exactly arrives at the result of Bonnor. 

Our equation (\ref{edeta13})  differs from the corresponding equation of Martel (see Eq. (8) in \cite{Martel}) in the case of the nonzero cosmological constant.

 The two descriptions differ in the part concerning the evolution of velocity perturbations. In the model of Bonnor    the  perpendicular velocity  components behave like   ${\vec V_T}\propto 1/a(t)$ \cite{Bonnor}; thus their length has to decrease. In the general relativistic analysis we have only a partly coincident behaviour of velocity perturbations --- $\partial_0(a \delta_U)\le 0$, assuming that $\partial_R\delta \ge 0$. In the case of dust-like perturbations --- with the vanishing speed of sound, $c^2_\mathrm{s}=0$ --- the velocity perturbation $\delta_U$ is strictly  decreasing. Positive velocity perturbations  might   decrease at least like the inverse of the scale factor, $1/a(t)$, but there is no a bound onto the absolute value of negative velocity disturbances $\delta_U$.   
 
\section{The influence of dark energy}
 
We shall  investigate how dark energy would influence the evolution of the  mass density contrast $\delta $ after the end of recombination epoch, that is for times
 $t\ge t_\mathrm{re}$. We neglect --- as in the whole paper ---  the contribution of  the radiation energy. The speed of sound $c_s$ is  negligible in this period and the evolution equation becomes 
\begin{equation}
\partial^2_0 \delta  -\frac{3}{2}H^2 \delta +2H \partial_0\delta = 0.
\label{edeta14}
\end{equation}
 
\subsubsection{Absence of dark energy }
In this case   the conformal factor  $a(t)\propto t^{2/3}$ and $H=2/(3t)$. The increasing solution of (\ref{edeta14}) reads $\delta (t)\propto a(t)\propto t^{2/3}$.  According to 
astronomical observations $a(t_0)/a(t_\mathrm{re})\approx 1100$ \cite{Weinberg08}; here $t_0$ is the present age of the Universe. Thus the  mass density contrast $\delta $ of dust-like perturbations of dust Friedman universes  would increase   1100 times since the end of the recombination era.

\subsubsection{Including  dark energy }
In this case  the coefficients --- $H(t)$ and $H^2(t)$ are given as related solutions of the Friedman equations (see Sec. III); the latter can be solved numerically, assuming dust and the cosmological constant. The evolution equation reads
 \begin{equation}
\partial^2_0 \delta  -\frac{3}{2}H^2 \delta +2H \partial_0\delta = 0.
\label{edeta14}
\end{equation} 
At the recombination era the material density $\varrho $ exceeds the dark energy density  $\varrho_\Lambda$ by a factor of the order of $10^8$. Thus as initial data we can choose 
\begin{equation}
\delta (t_\mathrm{re})=t_\mathrm{re}^{2/3}, ~~~~ \frac{d\delta  }{dt}|_{t_\mathrm{re}}=\frac{2}{3t_\mathrm{re}^{1/3}}
\label{initial}
\end{equation} 
--- these are data dictated by the solution $\delta (t)\propto a(t)$, valid in the case of no-dark energy.

The   solution of Eq. (\ref{edeta14}) with initial data (\ref{initial}) is very close to $\delta (t)=t^{2/3}$;  the difference becomes clear at relatively late times $t\ge t_0/10$ \cite{students}.

Assuming a flat universe with present data $\Omega (t_0)_\mathrm{d}=0.3$ and $\Omega_\mathrm{\Lambda}(t_0)=0.7$,
one gets $\delta (t_0)/\delta(t_\mathrm{re})\approx 975$ \cite{students}.
The cosmological constant  slows the process of formation of bound structures; its influence is comparable to that obtained from the equation of Martel --- see    \cite{ Weinberg08}.

\end{document}